\documentclass{iaus}
\usepackage{graphicx}

\title[$\tau$\,Sco: The discovery of the clones] 
{$\tau$\,Sco: The discovery of the clones}

\author[Petit et al.]   
{V. Petit$^1$, D.~L. Massa$^2$, W.~L.~F. Marcolino$^3$, G.~A. Wade$^4$, R. Ignace$^5$
 \and the MiMeS Collaboration}

\affiliation{$^1$Dept. of Geology \& Astronomy, West Chester University, West Chester, PA 19383, USA  \\ 
email: {\tt VPetit@wcupa.edu} \\[\affilskip]
$^2$Space Telescope Science Institute, 3700 N. San Martin Drive, Baltimore, MD 21218, USA \\
[\affilskip]
$^3$Observat\'orio Nacional-MCT, CEP 20921-400, S\~ao
Crist\'ov\~ao, Rio de Janeiro, Brasil  \\
[\affilskip]
$^4$Dept. of Physics, Royal Military College of Canada, Kingston, Canada, K7K 4B4 \\
[\affilskip]
$^5$Dept. of Physics \& Astronomy, East Tennessee State University, Johnson City, TN 37614, USA  \\
}

\pubyear{2010}
\volume{272}  
\pagerange{119--126}
\jname{Active OB stars: structure, evolution, mass loss, and critical limits}
\editors{C. Neiner, G. Wade, G. Meynet \& G. Peters, eds.}
\begin{document}

\maketitle

\begin{abstract}
The B0.2 V magnetic star $\tau$\,Sco stands out from the larger population of massive magnetic OB stars due to its remarkable, superionized wind, apparently related to its peculiar magnetic field - a field which is far more complex than the mostly-dipolar fields usually observed in magnetic OB stars. $\tau$\,Sco is therefore a puzzling outlier in the larger picture of stellar magnetism - a star that still defies interpretation in terms of a physically coherent model.

Recently, two early B-type stars were discovered as $\tau$\,Sco analogues, identified by the striking similarity of their UV spectra to that of $\tau$\,Sco, which was - until now - unique among OB stars. We present the recent detection of their magnetic fields by the MiMeS collaboration, reinforcing the connection between the presence of a magnetic field and a superionized wind. We will also present ongoing observational efforts undertaken to establish the precise magnetic topology, in order to provide additional constrains for existing models attempting to reproduce the unique wind structure of $\tau$\,Sco-like stars. 
\end{abstract}

\firstsection 
\section{The young magnetic B-type star $\tau$\,Sco}

The magnetic field of $\tau$\,Sco is unique because it is structurally far more complex than the mostly-dipolar 
fields ($l=1$) usually observed in magnetic OB stars, with significant power in spherical-harmonic modes 
up to $l=5$ and a mean surface field strength of $\sim300$\,G \cite[(Donati et al. 2006)]{Donati06}.
$\tau$\,Sco also stands out from the crowd of early-B stars because of its stellar wind anomalies, as diagnosed through its odd UV spectrum. These anomalies, unique to this star, are indicative of a highly ionised outflow. 

Interestingly, the wind lines of $\tau$\,Sco vary periodically with the star's 41 d rotation period \cite[(Donati et al. 2006)]{Donati06}. Clearly the magnetic field exerts an important influence on the wind dynamics. What is not clear is whether the wind-line anomalies described above are a consequence of the unusual complexity of $\tau$\,Sco's magnetic field, a general consequence of wind confinement in this class of star, or perhaps even unrelated to the presence of a magnetic field.

\section{The $\tau$\,Sco Clones}

We present two early B-type stars - HD\,66665 and HD\,63425 - that we identified to be the first $\tau$\,Sco analogues. These stars were first discovered by their UV spectra, which are strikingly similar to the UV spectrum of $\tau$\,Sco. 
Spectropolarimetric observations of HD\,66665 and HD\,63425 were taken with ESPaSOnS at the Canada-France-Hawaii Telescope, in the context of the Magnetism in Massive Stars Large Program. We acquired 4 high-resolution, broad-band, intensity (Stokes I) and circular polarisation (Stokes V) for each star. 

\begin{figure}
\begin{center}
 \includegraphics[width=2.25in]{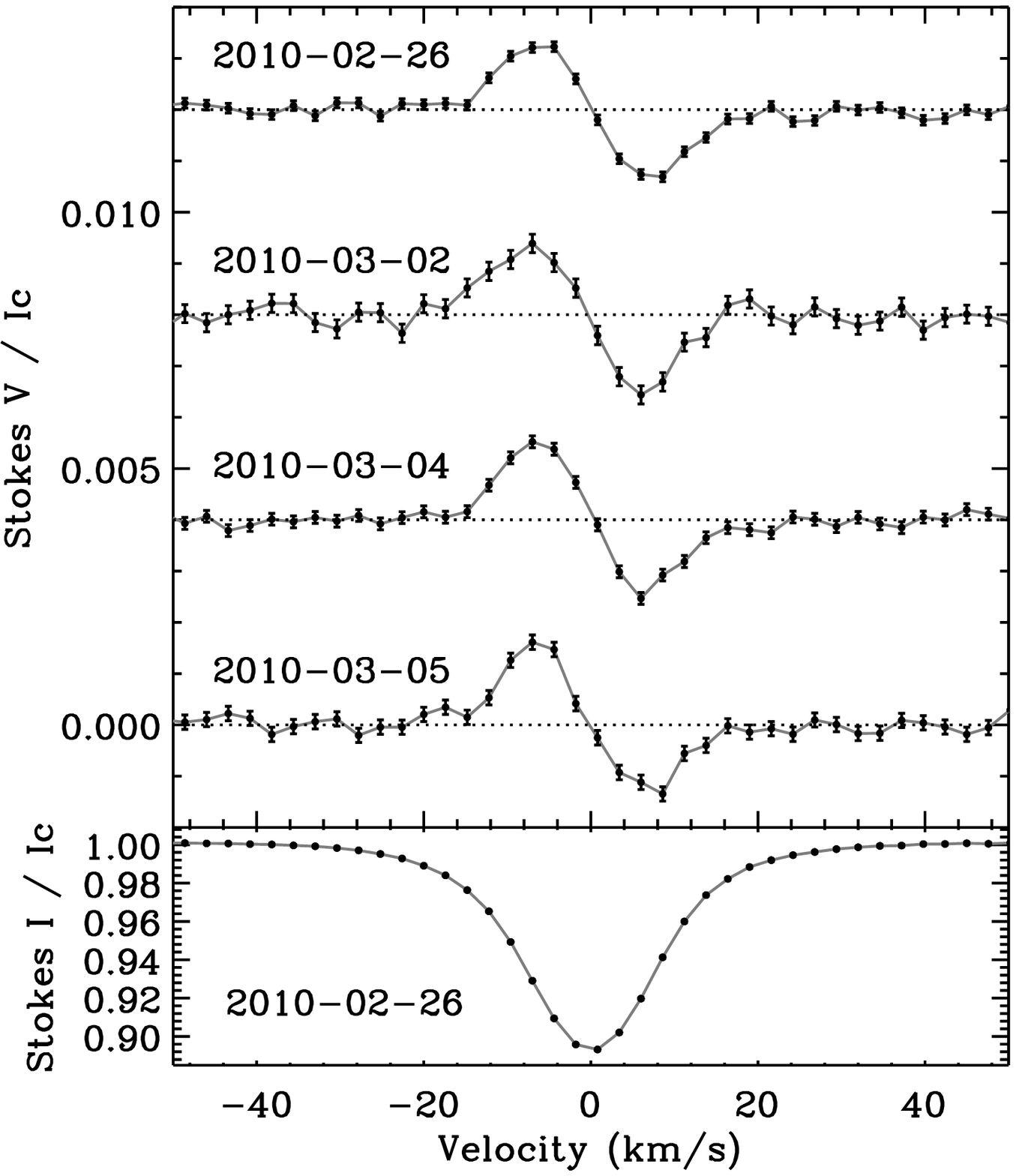} 
 \includegraphics[width=2.25in]{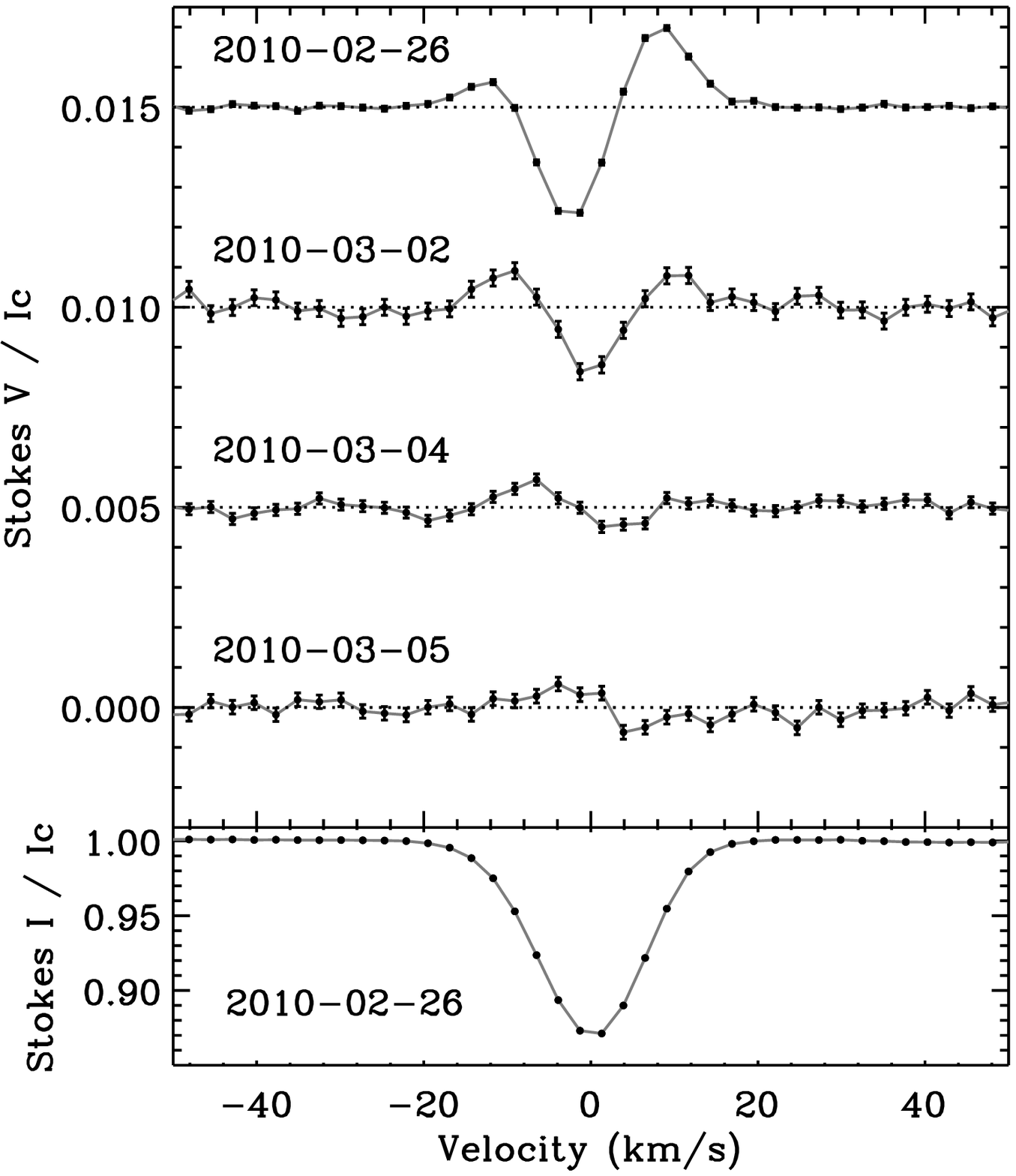} 
 \caption{Mean Stokes I absorption line profiles (bottom) and circular polarisation Stokes V profiles (top) of HD\,63425 (left) and HD\,66665 (right) obtained with ESPaDOnS.}
   \label{fig1}
\end{center}
\end{figure}

In order to determine the stellar and wind parameters of HD\,66665 and HD\,63425 we used non-LTE model 
atmospheres from the CMFGEN code \cite[(Hillier \& Miller 1998)]{Hillier98}. The physical parameters are similar to $\tau$\,Sco's. 
In order to increase the magnetic sensitivity of our data, we applied the Least-Squares Deconvolution (LSD) procedure \cite[(Donati et al. 1997)]{Donati97}, which enables the simultaneous use of many spectral lines to detect a magnetic field Stokes V signature. 
All observations led to a significant detection of a (time-variable) magnetic signal (see Figure \ref{fig1}). The same analysis was performed on the diagnostic null profiles, and no signal was detected. 

As the exact rotation phases of our observations are not known, we used the method described by \cite[Petit et 
al. (2008, in prep)]{Petit08}, which compares the observed Stokes V profiles to a rotation independent, dipolar 
oblique rotator model, in a Bayesian statistic framework. We can obtain a conservative estimate of the surface field strength, in a dipolar approximation.
The average surface field strengths are of the same scale as the surface field of $\tau$\,Sco. However, more 
phase-resolved observations are required in order to assess the potential complexity of their magnetic field, 
and verify if it is linked to the wind anomalies.

\end{document}